\documentclass[prd,12pt,nofootinbib,preprint,onecolumn,a4paper]{revtex4-1}
\pdfoutput=1
\usepackage[T1]{fontenc}
\usepackage[utf8x]{inputenc}
\usepackage{amsmath,amssymb}
\usepackage{epsfig}
\usepackage{url}
\usepackage{subfigure}
\usepackage{graphicx}
\usepackage{grffile}
\usepackage[usenames,dvipsnames]{color}
\usepackage{slashed}
\usepackage{array}
\usepackage{multirow}
\usepackage{xcolor}
\usepackage[colorlinks,citecolor=blue]{hyperref}
\usepackage{color}
\usepackage{cancel}
\usepackage{gensymb}

\def\bar {\overline}

\def\be {\begin{equation}}
\def\ee {\end{equation}}
\def\beq {\begin{equation}}
\def\eeq {\end{equation}}
\def\bea {\begin{eqnarray}}
\def\eea {\end{eqnarray}}

\newcommand{\besub}{\begin{subequations}}
\newcommand{\eesub}{\end{subequations}}

\def\beq{\begin{equation}}
\def\eeq{\end{equation}}
\def\barr{\begin{array}}
\def\earr{\end{array}}

\def\q2 {q^2}

\def\bt{\begin{table}}
\def\et{\end{table}}

\def\mET{\slashed{E}_T}

\begin{document}

%
%
%

\vskip 30pt 

\begin{flushright}
HRI-RECAPP-2022-007
\end{flushright}

\begin{center}
{\Large \bf Collider Signatures of a Scalar Leptoquark and
Vector-like Lepton in Light of Muon Anomaly} \\
\vspace*{1cm}
{\sf ~Nivedita Ghosh$^{a,}$\footnote{niveditaghosh@hri.res.in},
     ~Santosh Kumar Rai$^{a,}$\footnote{skrai@hri.res.in},
     ~Tousik Samui$^{b,}$\footnote{tousik.pdf@iiserkol.ac.in}}\\
\vspace{10pt}
{\small }
{ $^a$Regional Centre for Accelerator-based Particle Physics,\\
    Harish-Chandra Research Institute,
    A CI of Homi Bhabha National Institute,\\
    Chhatnag Road, Jhunsi, Prayagraj -- 211019, India\\
    $^b$Department of Physical Sciences,
    Indian Institute of Science Education and Research Kolkata,
    Mohanpur, 741246, India}
\end{center}

\begin{abstract}
Despite the immense success of the Standard Model, the hunt
for physics beyond the Standard Model has continued.
Extension of the SM gauge group or the particle content of
the SM remains a viable solution to many observed anomalies,
such as the anomalous magnetic moment of muon, flavor
violation, dark matter, etc. In this work, we consider a BSM
model which includes a leptoquark, a vector-like lepton, and a
real scalar singlet. The model accounts for a dark matter
candidate through a $\mathbb{Z}_2$ symmetry in addition to
predicting the muon $(g-2)$ in agreement with the
experimental measurement. We also show that the experimental
measurements of the lepton flavor violation are satisfied in
a wide range of parameter space. The viable parameter space
of the model is used to study the collider 
signatures arising from the pair production of the 
leptoquark in the $2\mu+2b+\mET$ channel at the 14~TeV LHC run.
\end{abstract}
\maketitle

\renewcommand{\thesection}{\Roman{section}}  
\setcounter{footnote}{0}  
\renewcommand{\thefootnote}{\arabic{footnote}}  

\section{Introduction}
The success of the Standard Model (SM) predictions of particle physics is astonishingly precise. 
All the predicted particles have been observed in various experiments. The last particle to
be discovered is the Higgs boson\,\cite{ATLAS:2012yve,CMS:2012qbp}. Despite its huge success, it has limitations in explaining several anomalies observed in precision measurements and other
experiments\,\cite{Muong-2:2015xgu,Muong-2:2021vma,MEG:2016leq,BaBar:2009hkt,LHCb:2014vgu,HFAG:2014ebo,FermilabLattice:2016ipl}.
These limitations include neutrino mass and mixing observations, dark matter (DM) in the universe, several
anomalies in the flavor sector, and $(g-2)_\mu$ anomaly, among many others. The quest for physics 
beyond the Standard Model (BSM) that could account for these observations has become a 
popular area of research in particle physics.

One particular anomaly, which has been of interest for a 
long time, is the gyromagnetic ratio of the electromagnetic interaction of leptons, especially for muon ($\mu$). 
There seems to be a $4.2\sigma$ discrepancy between the SM prediction and the
latest measurement at the E989 experiment at Fermi National
Laboratory (FNAL)\,\cite{Muong-2:2021vma}. This discrepancy
was at 3.7$\sigma$ in the earlier observation at the
Brookhaven National Laboratory
(BNL)\,\cite{Muong-2:2006rrc}. The solution to this long-term discrepancy between the SM prediction and the experimentally observed value exist in many new physics (NP) scenarios. 
The most popular ones invoke new particles and/or symmetries 
that provide additional contributions to the SM prediction. There are, for example, the popular non-supersymmetric models 
with single-field extensions are $Z'$ models\,\cite{Chiu:2014oma,Bauer:2015knc,ColuccioLeskow:2016dox,Datta:2019bzu,Marzocca:2021azj,Zhang:2021dgl,Ko:2021lpx,Cheng:2021okr},
singlet-scalar
extension\,\cite{Gninenko:2016ziu,Liu:2020qgx,Saez:2021qta,Capdevilla:2021kcf},
vector-like lepton (VLLs)\,\cite{Dermisek:2013gta,Poh:2017tfo,Crivellin:2018qmi,DeJesus:2020yqx,Abdallah:2020biq,Frank:2020smf,Chun:2020uzw,Chakrabarty:2020jro,Crivellin:2021rbq},
or extensions of the Higgs doublet to
two\,\cite{Ghosh:2020tfq,Abdallah:2020vgg,Ghosh:2021jeg,Hernandez:2021tii,Dermisek:2021ajd,Hernandez:2021iss,Arcadi:2021yyr}
or more doublets. While these extensions might give rise to
enough contribution to the anomalous magnetic moment of
$\mu$, they also contribute to the lepton flavor violation.
Since we do not have any evidence for any lepton flavor
violation (LFV), simultaneously achieving both goals
in a simplistic and universal model is difficult. 
The usual way around this is to consider non-universal couplings 
of the new particles with the SM particles. There are models with two-particle extensions where it is relatively easy to explain
muon anomaly and non-observation of LFV at the same
time\,\cite{Freitas:2014pua,Kowalska:2017iqv,Calibbi:2018rzv,Athron:2021iuf}.
The most popular two-particle extensions usually consider
a combination of scalars and fermions. For the lepton sector
analysis, the fermion is usually a VLL. In addition to 
such extensions, other ideas also have possible explanations for the $(g-2)_\mu$ anomaly. These ideas mainly include extra-dimensional
models\,\cite{Das:2004ay,Moch:2014ofa},
technicolor models\,\cite{Doff:2015nru}, composite
models\,\cite{Das:2001it,Hong:2016uou,Cline:2017aed,Xu:2022one}, leptoquark models\,\cite{Crivellin:2019dwb,Bhaskar:2022vgk}, etc. In
this work, we attempt to explain the $(g-2)_\mu$ anomaly
while allowing solutions to some of the observed flavor
anomalies by extending the SM particle content. In our
scenario, we have a scalar leptoquark, a real scalar 
singlet, and a pair of VLLs added to the
SM\,\cite{Arnan:2016cpy,Dhargyal:2018bbc}. The leptoquark
and singlet scalar usually contribute to the magnetic
moment of muon via the chiral breaking terms which are
proportional to the mass of the heavy particle running in
the loop. The VLL is responsible in giving necessary
couplings to the $\mu, \, \tau$, and $e$, so that we get
significant contribution to the $(g-2)_\mu$, 
while the LFV does not receive any large contributions. In our
scenario, due to an explicit $\mathbb{Z}_2$ symmetry the
scalar leptoquark is instrumental in only addressing the
meson flavor anomalies but the VLL and singlet scalars
contribute to muon anomaly as well as LFV at one-loop.
This model was considered before for explaining
various flavor anomalies, for example,
$R_D^{(*)}$, $b \to s \mu \mu$, $B_s-\bar{B_{s}}$
oscillation\,\cite{Dhargyal:2018bbc}. In addition, the
presence of a neutral scalar or the lightest neutral
component of the VLL also explains the dark
matter puzzle where the neutral scalar or fermion play the
role of dark matter. The presence of a leptoquark and charged 
VLL in the model also lead to interesting collider signals. We 
note that most models (unlike ours), introduce a leptoquark 
but do not augment it with an odd $\mathbb{Z}_2$ parity. 
The absence of the $\mathbb{Z}_2$ parity allows the leptoquark
to couple to SM particles singly, which leads to its direct decay to
SM final states. These give strong bounds on the
leptoquark mass that can be produced via strong interactions at
hadron colliders and then reconstructed through the SM decay
products\,\cite{ATLAS:2020dsk,D0:2006wsn,ATLAS:2016wab,Okumura:2021yjv}. In our case,
the production of the leptoquark followed by its decay to
VLL and quarks lead to a different signal when compared to the
studies available in the literature. The VLL then decays to
charged SM leptons and a neutral singlet scalar. The decays finally 
lead us to a signal of dimuon plus dijet with missing transverse
energy. We study the viability of observing the signal  at the current and future runs of the 
Large Hadron Collider (LHC) in this work.

The paper is organized as follows. In Section~\ref{Model},
we describe the model. In Section~\ref{Muon},
we discuss the allowed parameter space of our model that explain the muon anomalous magnetic moment 
and lepton flavor violations. We discuss the constraints from electroweak precision (EW) measurements in 
Section~\ref{EWP}. In Section~\ref{QFV}, we examine the constraints coming from quark flavor-violating
processes. We finally look for distinctive collider signals at LHC and discuss our results in 
Section~\ref{coll} and conclude in Section~\ref{conc}.
\newpage

\section{Model}~\label{Model}
We propose a new physics model that may resolve several flavor anomalies observed in recent experiments. 
The model extends the SM particle content by adding new particles that include a leptoquark ($\Phi$), a pair of
$SU(2)$ doublets ($L_{4L},\ L_{4R}$), and a real scalar singlet field ($S$). We augment the SM symmetry with an
additional discrete  $\mathbb{Z}_2$ symmetry under which all the SM
states are even while the new states are odd. The charge of the new fields under the 
$\mathcal{G}=SU(3)_C \times SU(2)_L \times U(1)_Y \times \mathbb{Z}_2$ is tabulated in Table~\ref{tab:charges}.
\begin{table}[h]
\begin{tabular}{|c|c|c|c|c|}
\hline
~~Particles~~ & ~~$SU(3)_C$~~ & ~~$SU(2)_L$~~ & ~~$U(1)_Y$~~ & ~~$\mathbb{Z}_2$~~ \\
\hline
$\Phi$ & 3 & 1 & 2/3 & $-1$\\
\hline
$L_{4L}$ & 1 & 2 & $-1/2$ & $-1$\\
\hline 
$L_{4R}$ & 1 & 2 & $-1/2$ & $-1$\\
\hline
$S$ & 1 & 1 & 0 & $-1$\\
\hline
\end{tabular}
\caption{New fields and their charges.}
\label{tab:charges}
\end{table}

The gauge invariant Lagrangian for the BSM sector can be written as 
\begin{eqnarray}
\mathcal{L} &\supset& -M_\Phi^2 \Phi^\dagger \Phi - M_S^2 S^2 -\lambda_{H\Phi} H^\dagger H \Phi^\dagger \Phi - \lambda_{S\Phi} \Phi^\dagger \Phi S^2 - \lambda_{HS} H^\dagger H S^2 - \lambda_\Phi \left(\Phi^\dagger \Phi\right) - \lambda_S S^4 \nonumber\\
            & & -\left\{h_i \bar L_{4R} Q_{Li} \Phi^\dagger + h'_j \bar L_{4R} L_{Lj} S + M_F \bar L_{4L} L_{4R} + h.c.\right\},
            \label{Lag}
\end{eqnarray}
where $H$ is the SM Higgs doublet, $Q_{Li}$ and $L_{Lj}$,
($i,j=1,2,3$) are the SM quark and lepton doublets,
respectively. The VLL doublet $L_{4} = (\nu_{4},
\ell_{4}^-)^T$ has $\nu_4$ and $\ell^-_4$ as the neutral
and charged components, respectively. Since all the new particles are odd and SM particles are 
even under the unbroken $\mathbb{Z}_2$ symmetry, the odd particles do not mix with the even particles. 
This prevents any modification to the couplings of the particles in the SM sector as the unbroken $\mathbb{Z}_2$
prevents the scalar $S$ from acquiring a vacuum expectation
value (VEV). 
However, the presence of $\lambda_{H\Phi}$ and
$\lambda_{HS}$ terms lead to modified mass for $\Phi$ and $S$ 
which shift from $M_\Phi$ and $\sqrt{2}M_S$ after electroweak
symmetry breaking. 
The absence of mixing between the SM Higgs boson and $S$ keeps the couplings of SM particles with 
the SM Higgs boson unaffected. 
The new scalars in the model are taken to be heavier than the SM Higgs boson such that we do not have any additional 
decay modes of the SM Higgs boson with respect to the SM. However, the effect of these new heavy states which are
both colored and electrically charged would appear through loops. This would alter the
$hgg$ and $h\gamma\gamma$ effective couplings, which appear at the one-loop level.
We further discuss this in Section~\ref{coll}.

We note that almost all the properties of the SM scalar sector remain unaffected at the tree-level in this model. There 
would be corrections to SM interactions at the loop-level, giving the  necessary contributions to the
muon $(g-2)$ that could explain the anomaly. The terms in the Lagrangian with the Yukawa couplings 
$h_i$ and $h'_j$ are responsible for these contributions, where  $i,j=1,2,3$ represent the generation indices.
Note that the Yukawa term in Eq.~\ref{Lag} containing $h_i$ is written 
in terms of the flavor eigenbasis of quarks which finally mix via the Cabibbo-Kobayashi-Maskawa (CKM) mixing. 
After CKM mixing, the couplings of SM quarks in the physical (mass) eigenbasis become
\begin{equation}
h_i^\text{ph} \to \sum_{j=1}^{3}h_j U_{ji}^d,
\end{equation}
where $U^d$ is the mixing matrix of down-type quarks. The new Yukawa couplings would also give rise to 
new contributions to several other phenomena. For example, the presence of the $h_i$ term in the Lagrangian 
leads to additional contributions to the $K^0-\bar{K^0}$ and $B^0-\bar{B^0}$ oscillations. These two oscillation measurements agree well with the SM prediction. Therefore any new physics contribution to the $K^0-\bar{K^0}$ and 
$B^0-\bar{B^0}$ oscillations must be small and will help in constraining the parameters of the BSM model. 
In our model this is achieved by choosing leptoquark coupling to the first generation quark to be negligible, {\it i.e.} $h_{1,2}^\text{ph} \simeq 0$\,\cite{Dhargyal:2018bbc}. At the same time, there is still room for new physics in the $B_s^0-\bar{B_s^0}$ oscillation data which gives a constraint $|h_2^\text{ph} h_3^\text{ph}| \lesssim $ 0.65\,\cite{Dhargyal:2018bbc,DiLuzio:2017fdq,ParticleDataGroup:2020ssz}.

The lepton sector is affected by the term containing $h_j'$
coupling. Instead of leptoquark, we have a real scalar
involved in this coupling. The enhancement in the
$(g-2)_\mu$ can be achieved by considering VLL and the real
scalar in the loop. At the same time, the lepton flavor
violating decay like $\mu\to e \gamma$ should not yield a
large value. This can be achieved by taking large values for
the $h'_2$ and keeping other $h'$ negligible.
Moreover, if the real scalar is the lightest among all the
new states, it may play the role of DM in this model. The
possibility of explaining the DM in this model will be
discussed in the later part of this article. In this work,
we mainly focus on the muon anomaly, lepton and quark flavor
violation, and the signatures of the leptoquark at the LHC. 

\section{Muon Anomaly and Lepton Flavor Violation}\label{Muon}
The gyromagnetic ratio of muon ($g_{\mu}$) is exactly 2 at the tree-level. However quantum corrections are induced through 
contributions from higher-order loops within the SM and the deviation from its tree-level value is denoted by
$a_{\mu}=\dfrac{g_{\mu}-2}{2}$, known as the anomalous magnetic moment of the muon. In the SM, the current value
reads\,\cite{Aoyama:2020ynm,Aoyama:2012wk,Aoyama:2019ryr,Czarnecki:2002nt,Gnendiger:2013pva,Davier:2017zfy,Keshavarzi:2018mgv,Colangelo:2018mtw,Hoferichter:2019mqg,Davier:2019can,Keshavarzi:2019abf,Kurz:2014wya,Melnikov:2003xd,Masjuan:2017tvw,Colangelo:2017fiz,Hoferichter:2018kwz,Gerardin:2019vio,Bijnens:2019ghy,Colangelo:2019uex,Blum:2019ugy,Colangelo:2014qya}
\begin{equation}
 a^\text{SM}_{\mu} = 116591810(43) \times 10^{-11}.
\end{equation}

The recent results from the {\sc ``Muon~G-2''} at
Fermilab\,\cite{Muong-2:2015xgu} from their first run data
provides the anomalous magnetic moment to
be\,\cite{Muong-2:2021vma}
\begin{equation}
 a^\text{exp−FNAL}_{\mu} = 116592040(54) \times 10^{-11}.
 \label{Fermi}
\end{equation}

The combined new world average (combination of recent FNAL\,\cite{Muong-2:2021vma} and older BNL(2006)\,\cite{Muong-2:2006rrc} data) is published
as\,\cite{Muong-2:2021ojo}
\begin{equation}
 a^\text{exp−comb}_{\mu} = 116592061(41) \times 10^{-11}.
\end{equation}

The difference between the experimental observation and the
SM prediction, defined as $\Delta a_{\mu}$, amounts to a
4.2$\sigma$ discrepancy, which provokes one to look beyond the
SM.
\begin{equation}
\Delta a_{\mu} = a^\text{exp−comb}_{\mu} - a^{SM}_{\mu} = 251(59) \times 10^{-11}.
\end{equation}

\begin{figure}[htpb!]{\centering
\includegraphics[height = 5.5 cm, width = 8 cm]{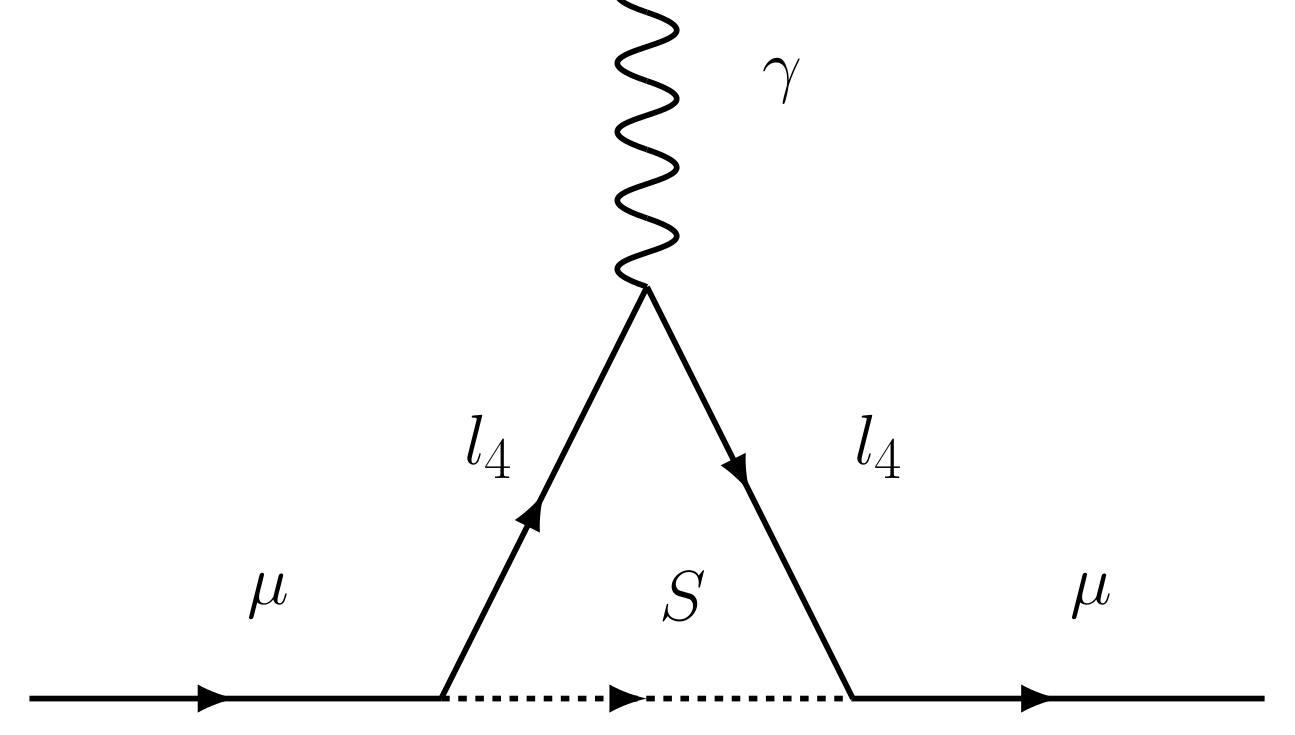}}
\caption{Feynman diagram for the NP contribution to
$(g-2)_\mu$ at one-loop.} \label{Feyn}
\end{figure}

The  $a_{\mu}$ is chirality flipping and is generated in our model by the Feynman diagram corresponding to the 
new physics (NP) contribution at one-loop as shown in Fig.~\ref{Feyn}. The contribution comes from the scalar 
$S$ and the VLL in the loop and the expression for $\Delta a_\mu$ is given by\,\cite{Arnan:2016cpy}
\begin{eqnarray}
\Delta a_\mu = \frac{m_\mu^2 |h'_2|^2}{8\pi^2 M_{\ell_4}^2} f\left(\frac{M_S^2}{M_{\ell_4}^2}\right),
\end{eqnarray}
where $m_\mu$ is the mass of muon and 
\begin{eqnarray}
f(x) = \frac{1-6x-6x^2\ln x +3x^2 + 2x^3}{12(1-x)^4}.
\end{eqnarray}

\begin{figure}[htpb!]{\centering
\subfigure[]{
\includegraphics[height = 5.5 cm, width = 8 cm]{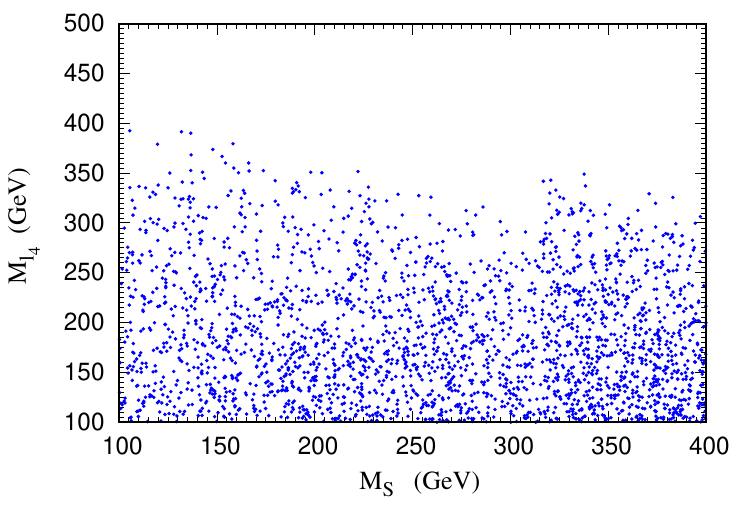}}
\subfigure[]{
\includegraphics[height = 5.5 cm, width = 8 cm]{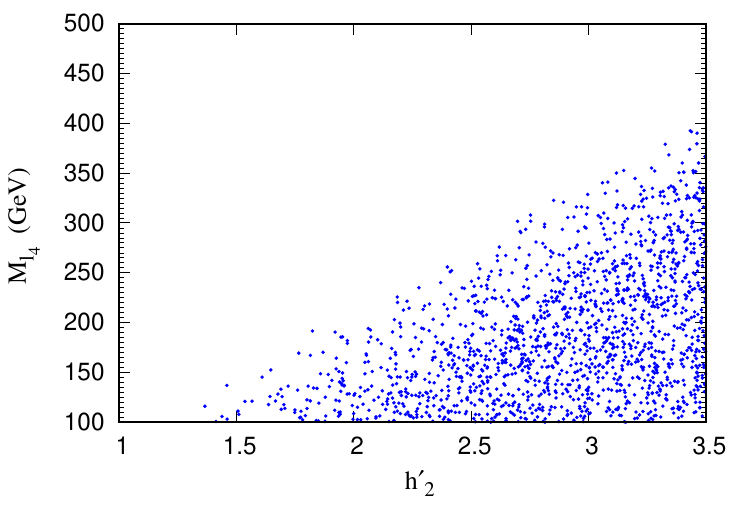}}
}
\vspace{-5mm}
\caption{The allowed parameter space satisfying muon anomaly
in (a) $M_S$-$M_{\ell_4}$ and (b) $h'_2$-$M_{\ell_4}$ plane.
For these plots, we have varied the parameter as given in
Eq.~(\ref{scanrange}).}
\label{muanomaly}
\vspace{-3mm}
\end{figure}
 
We show the allowed range of the parameter space in Fig.~\ref{muanomaly}
which satisfies the experimental value of $\Delta a_\mu$ within a
$3\sigma$ range. The parameters in our model are scanned over: 
\begin{eqnarray}
 M_{\ell_4} \in [102.6:500]~\text{GeV},\qquad M_S \in [100:400]~\text{GeV},\qquad h'_2 \in [1:3.5] \label{scanrange}
\end{eqnarray}
 to highlight the regions of parameter space which can explain the $(g-2)_\mu$ observation.
The mass of the charged lepton $M_{\ell_4}$ has a lower
bound of 102.6~GeV from Large Electron Positron
collider (LEP)\,\cite{ParticleDataGroup:2020ssz} and hence, in the
scan, the lower range of $M_{\ell_4}$ has been set to this
value. From Fig.~\ref{muanomaly}(a), we see that $M_{\ell_4}
\lesssim 400$ GeV is preferred to satisfy the muon
anomaly data. Larger values of $M_{\ell_4}$ would render the Yukawa coupling $h'_2$ non-perturbative. 
The allowed values for the scalar mass can be, however, equal, heavier or lighter than the VLL mass. 
In our model, we preferably set the scalar $S$ to be the DM candidate, which sets $M_S < M_{\ell_4}$. 
It is worth pointing out that the leptoquark mass $M_\Phi$ does not play any role 
which is expected, as the leptoquark which is odd under $\mathbb{Z}_2$ will not contribute at one-loop. 
The anomaly prefers a large value for the Yukawa coupling as one can see from Fig.~\ref{muanomaly}(b).  We
find that $h'_2 \gtrsim 1.5$ are more favorable for all mass values of the VLL. The parameters that 
explain the muon anomaly are chosen for our collider analysis which we discuss later in Section~\ref{coll}. 

We note that similar diagrams as the muon anomaly diagram, with external muons replaced by appropriate leptons, will contribute to the LFV decay modes. Though the violation of lepton flavor
has been observed in neutrino oscillation\,\cite{Super-Kamiokande:1998kpq,SNO:2002tuh}, non-observation 
of any significant LFV in the charged lepton sector put strong constraints on LFV processes. 
The strongest bound in the $\mu$--$e$ sector 
(BR$(\mu \to e\gamma) < 4.2\times10^{-13}$) comes from the MEG
experiment\,\cite{MEG:2016leq}. Similar to the $\mu$--$e$ sector sector, we also get constraints 
from $(\tau \to e\gamma)$ and $(\tau \to \mu\gamma)$ decay branching ratios (BR).
The current bound on these lepton flavor conversions are~\cite{BaBar:2009hkt} 
\begin{align*}
{\rm BR} (\tau \to e\gamma) < 3.3\times10^{-8} \, ,&&  {\rm BR} (\tau \to \mu\gamma) < 4.4\times10^{-8} \,. 
\end{align*}
As pointed out before, these constraints can be avoided easily in our model by choosing $h'_1$ and $h'_3$ small. 
We choose values for these parameters in the following range: 
$h'_1 \in [10^{-5}\,:\,10^{-4}]$, and, $h'_3 \in [0.01\,:\,0.1]$ which are allowed by the above LFV constraints. 

\section{Electroweak Precision Measurement}~\label{EWP}
\begin{figure}[!h]
\includegraphics[width=0.55\textwidth]{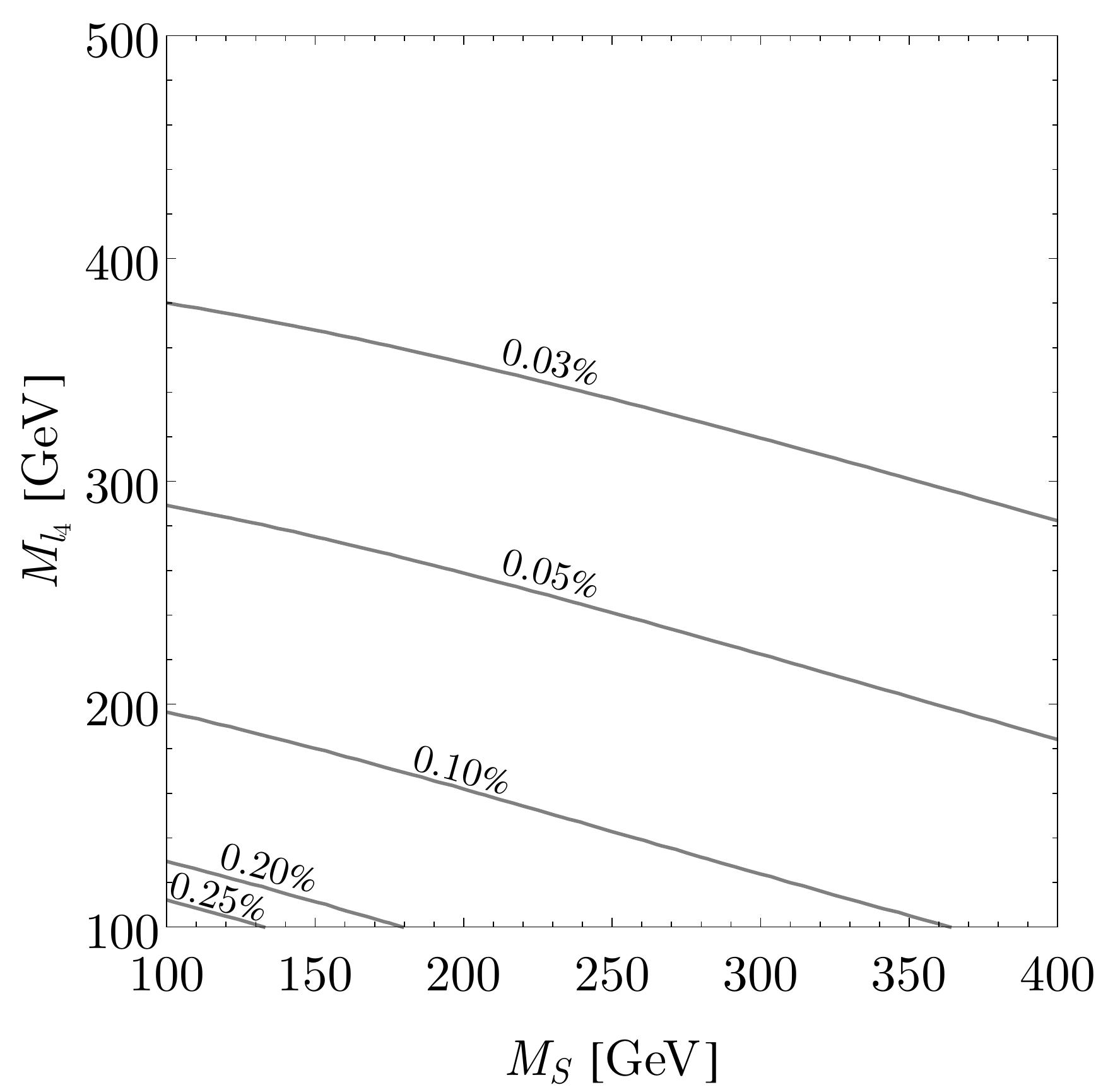}
\vspace{-3mm}
\caption{Contour plot of $\big|\delta g_{L}^\mu/g_{L,\text{SM}}^\mu(q^2=M_Z^2)\big|$ as a function of $M_S$ and $M_{l_4}$ with $h_2'=3.0$.  }
\label{fig:ewp}
\vspace{-8mm}
\end{figure}

We note that the observed anomalous magnetic moment of the muon and $R_K^{(*)}$ anomaly  prefers a relatively large value of $h_2'$ in our model. A large Yukawa coupling could have several unwanted consequences as it can lead to large contributions to various 
subprocesses within the SM, even when the new physics particles appear in the loops. We already discussed how our model 
avoids LFV constraints but  an immediate concern arises from electroweak precision measurements at the Large Electron-Positron Collider (LEP). The choice of large values for $h_2'$ can not only cause corrections for the muon mass but also alter the SM $Z\mu^+\mu^-$ coupling, which was precisely determined at LEP. In our model,  the additional contribution to the interaction vertex of 
$Z\mu^+\mu^-$ comes from a similar one-loop diagram shown in Figure~\ref{Feyn}. The relative change of $Z\mu^+\mu^-$ coupling with respect to its SM value can be expressed as
\begin{eqnarray}
\frac{\delta g_{L}^\mu}{g_{L,\text{SM}}^\mu}(q^2) = \frac{q^2}{32\pi^2 M_{\ell_4}^2} |h_2'|^2 \,G\left(\frac{M_S^2}{M_{\ell_4}^2}\right)
\end{eqnarray} 
where $q$ is the momenta carried by the $Z$ boson and the loop function
\begin{eqnarray}
G(x) = \frac{7-36x+45x^2-16x^3+(12x^3-18x^2)\log\,x}{36(x-1)^4}
\end{eqnarray}
From the LEP, the upper limit for the absolute value of $\delta g_{L}^\mu/g_{L,\text{SM}}^\mu(q^2=M_Z^2)$ is 0.8\%\,\cite{Arnan:2016cpy}. For our model, we plotted a contour plot of $\big|\delta g_{L}^\mu/g_{L,\text{SM}}^\mu(q^2=M_Z^2)\big|$ as function of $M_S$ and $M_{l_4}$ in Figure~\ref{fig:ewp} with $h_2'=3.0$. The number on the contour lines represent the value of $\big|\delta g_{L}^\mu/g_{L,\text{SM}}^\mu(q^2=M_Z^2)\big|$ in our model. As can be seen from Figure~\ref{fig:ewp}, within the range of the contour plot, the relative change of $Z\mu^+\mu^-$ coupling due to new physics is always less than $0.3\%$. Since we are interested in the mass range $M_S>100$~GeV and $M_{l_4}>100$~GeV, we are safe from the LEP constraints in the range of parameters which we shall use for our LHC analysis that can simultaneously explain the muon anomaly. 

\section{Quark Flavor Violation}~\label{QFV}
The presence of leptoquarks and their interactions with SM quarks and leptons 
can also induce flavor-violating decays of hadrons. As pointed out earlier, the terms containing 
$h_i$ and $h'_i$ couplings induce extra contributions to flavor violation
in the quark sector via loop. In this work, we consider the one-loop contribution to the quark flavor
violation (QFV), primarily focusing on the constraints coming from $b\to s$ transition, 
observed in the decays of $B$ meson. The quark level transitions in such decays arise primarily from three modes,
viz. (i) $b \to s \ell^+ \ell^-$, (ii) $b\to s \bar\nu \nu$
and (iii) $b\to s\gamma$. In addition to $B$ meson decays, the $B_s$-$\bar{B}_s$ mixing also provides 
the information for $b\to s$ transition. 

We first provide a brief account of the recent experimental results and SM predictions for $b\to s$ transition. 
The most recent observation for $b\to s \ell^+ \ell^-$ transition comes from the measurement of the ratio
\begin{align*}
R_{K^{(*)}} = \frac{\text{BR}(B\to K^{(*)} \mu^+ \mu^-)}{\text{BR} (B\to K^{(*)} e^+ e^-)} 
\end{align*}
by LHCb\,\cite{LHCb:2014vgu}. The
measurement is in tension with SM and stands $2.6\sigma$ away from the SM prediction\,\cite{Bordone:2016gaq}. 
The $b\to s \bar\nu\nu$ transition is constrained from 
$B\to K^{(*)}\bar\nu \nu$ transition\cite{Buras:2014fpa} and our model gives no additional contribution to this mode.
The meson decay $B\to X_s \gamma$ provides constraints on $b\to s\gamma$ transition. 
The current experimental result and SM prediction for this decay probability
$\text{BR}(b\to s \gamma)$ are $(3.43 \pm 0.21 \pm 0.07)
\times 10^{-4}$\,\cite{HFAG:2014ebo} and $(3.36 \pm 0.23)
\times 10^{-4}$\,\cite{Misiak:2015xwa}, respectively. In the $B_s$-$\bar B_s$ system, the experimental result for 
$\Delta M_{B_s}$ is $1\sigma$ below the SM prediction\,\cite{FermilabLattice:2016ipl}.

From the theory side, the relevant NP contributions to $b\to s \mu^+\mu^-$ come from the following effective
Hamiltonian
\begin{eqnarray}
\mathcal{H} \supset -\frac{\alpha_\text{em} G_F}{\sqrt 2 \pi} V_{tb} V_{ts}^* (C_9\,\mathcal{O}_9 + C_{10}\mathcal{O}_{10})
\end{eqnarray}
where $\alpha_\text{em}$ and $G_F$ are fine-structure
constant and Fermi constant, respectively. $V_{tb}$ and
$V_{ts}$ are CKM matrix elements and $C_9$ and $C_{10}$ are
the Wilson coefficients of the operators
\begin{eqnarray}
\mathcal{O}_9 = \left(\bar s \gamma^\alpha P_L b\right) \left(\bar\mu \gamma_\alpha \mu\right), \qquad\qquad \mathcal{O}_{10} = \left(\bar s \gamma^\alpha P_L b\right) \left(\bar\mu \gamma_\alpha\gamma_5 \mu\right).
\end{eqnarray}

For $b\to s\gamma$, the NP contribution comes from the
effective Hamiltonian
\begin{eqnarray}
	\mathcal{H} \supset -\frac{m_b G_F}{4\sqrt 2 \pi^2} V_{tb} V_{ts}^* (C_7\,\mathcal{O}_7 + C_8\mathcal{O}_8)
\end{eqnarray}
where $m_b$ is the mass of bottom quark and $C_7$ and $C_8$
are Wilson coefficients corresponding to the operators
\begin{eqnarray}
\mathcal{O}_7 = e \left(\bar s \sigma^{\mu\nu} P_R b\right) F_{\mu\nu}, \qquad\qquad \mathcal{O}_8 = g_s \left(\bar s \sigma^{\mu\nu} T^a P_R b\right) G^a_{\mu\nu}.
\end{eqnarray}
For the $B_s$-$\bar B_s$ mixing, the effective Hamiltonian
is given by
\begin{eqnarray}
\mathcal{H} \supset C_{B\bar B} \left(\bar s \gamma^\mu P_L b\right) \left(\bar s \gamma_\mu P_L b\right).
\end{eqnarray}
The constraints coming from experimental measurements can
then be translated to give constraints on the above Wilson
coefficients. The combined fit provides the following bounds
on the Wilson coefficients\,\cite{Arnan:2016cpy}.
\begin{eqnarray}
&-1.14 \leq C_9 = -C_{10} \leq -0.23 &\qquad(3\sigma), \nonumber\\
&-2.8 \leq C_{B\bar B} \times (10^5\,\text{TeV}^2) \leq 1.3 &\qquad(3\sigma),\label{eqn:CBB}\\
& -0.098 \leq C_7 + 0.24\,C_8 \leq 0.070 &\qquad(2\sigma).\nonumber
\end{eqnarray}
For this model, the NP contribution to these Wilson coefficients can be obtained using the expressions given in Ref.\,\cite{Arnan:2016cpy}. One, of course, needs to ensure that the Wilson coefficients satisfy the bounds given in Eq.~(\ref{eqn:CBB}). This can be ensured by choosing appropriate values for $h_2$ and $h_3$ couplings which will be briefly discussed in the next section. 

\section{Collider Searches}~\label{coll}
In this section, we look at the collider signatures in our model. The most interesting signal comes 
from the leptoquark production. Since the leptoquark is $\mathbb{Z}_2$ odd its signals will not only entail 
a baryon and lepton number violation but also give large missing transverse momentum due to the presence of a 
final $\mathbb{Z}_2$ odd state in its decay cascade, which will go undetected. For this study we shall focus on the 
parameter space which can account for both muon anomaly and satisfy constraints coming from lepton and quark flavor violation. 

We consider the pair production process of the leptoquark. 
\begin{equation}
 p p \to \Phi\,\bar\Phi 
\end{equation}
The leptoquark carries charge $Q_{em} = +2/3$ and therefore can decay to both up-and down-type quark flavor. 
A few interesting points about the leptoquark in our model is worth highlighting before we discuss the signals arising 
from its production at LHC. Due to the unbroken $\mathbb{Z}_2$ parity, the leptoquark is forced to decay to a SM quark
and the new exotic lepton. This leads to an interesting possibility where its decay may not remain within the same fermion 
generation. This is because the VLL ($L_4$) couples more strongly to muon and the scalar $S$ compared to the other 
SM leptons. This enabled our model to account for the muon anomaly as discussed earlier. Note that the leptoquark 
which couples mostly to the third generation of SM quarks are weakly constrained compared to the leptoquarks which 
couple to the first two generations. We therefore consider the leptoquarks which have larger coupling to third generation 
SM quark and the exotic lepton. In such a case, and if kinematically allowed, the decay modes of the leptoquark 
will be as follows:
\begin{eqnarray*}
	\Phi &\to& b \, \ell_4^+ \to b\,\mu^+\, S\\
	     &\to& t \, \bar\nu_4 \to b\,W^+\,\nu S
\end{eqnarray*} 
As a multilepton final state at LHC will be much more clean,  we will consider only the leptonic decay of the $W$ boson\footnote{Hadronic decays of the $W$ boson will provide lower sensitivity due to the presence of huge QCD background.}. 
The collider signature of our interest then becomes a final state given by $2\mu+2b+\mET$. The SM subprocesses that can give 
rise to the same final state are dominated by the inclusive $t\bar{t}(+1 jet)$ with at least one extra hard jet\footnote{To generate this particular background, we have generated the $p p \to t \bar{t}$ and $p p \to t \bar{t} j$ in {\tt MADGRAPH}\,\cite{Alwall:2014hca} and used MLM matching scheme for the additional jet.}, and subdominant contributions come from the 
$t\bar{t}h$, $t\bar{t}V$, $VV$ and $VVV$, where $V=W^{\pm},Z$. 
In principle, the di-boson plus di-jets can also be a source of the background if both the light jets are mistagged 
as $b-jet$. However, the mistagging efficiency of a light jet identified as a $b-$jet is $\approx 1\%$ for $u,d,s$ and 
$\approx 10\%$ for the $c$ quark. In addition the production cross section is much smaller than the inclusive 
$t\bar{t}(+1 jet)$ background. We also note that the requirement of a high $p_T$ $b$-jet removes almost all of this 
background. For our analysis, we have chosen four benchmark points. We list them and 
the corresponding pair-production cross sections for the leptoquark in Table~\ref{bp}. For all the benchmark points, we 
have fixed the $h_i$ and $h'_i$ couplings which are shown in Table~\ref{tab:yuk}. 
We have ensured that the benchmark points satisfy the bounds discussed in Sections~\ref{Muon}, \ref{EWP}, and \ref{QFV}. For the QFV bounds, the values of the Wilson coefficients have also been tabulated in Table~\ref{bp}. These values are within the bounds provided in Eq.~(\ref{eqn:CBB}).

For the scalar sector, the modification of Higgs boson couplings is measured
experimentally in terms of Higgs boson signal strength, which is defined, in
the $\gamma\gamma$ channel, as
\begin{eqnarray}
\mu_{\gamma\gamma}= \frac{\left[\sigma(pp\to h)\times \text{BR}({h\rightarrow\gamma\gamma})\right]_\text{exp}}{\left[\sigma(pp\to h) \times \text{BR}({h\rightarrow\gamma\gamma})\right]_\text{SM}},
\label{eq:higgs_signal}
\end{eqnarray}
where $\sigma(pp\to h)$ is the production cross section of the Higgs and BR
denotes the branching ratio of the Higgs boson decaying to two photons. In our
model, we have calculated the values of the $\mu_{\gamma\gamma}$ using the
effective $hgg$ and $h\gamma\gamma$
couplings\,\cite{LHCHiggs:2013rie,LHCHiggs:2016ypw,ATLAS:2021vrm}. For the four benchmark points,
we provide these values in Table~\ref{bp}.
These values lie well within the 2$\sigma$ range of the CMS data $1.1 \pm 0.08$\,\cite{CMS:2022dwd}.

\begin{table}[!h]
\centering
\resizebox{\textwidth}{!}{
\begin{tabular}{|c|c|c|c|c|c|c|c|c|}
\hline
& $M_{\Phi}$ (GeV) & $M_{\ell_4}$ (GeV) & $M_S$ (GeV) &  $\sigma_{LO}(pp\to\Phi\bar\Phi)$ (fb) & $C_9$($-C_{10}$) & $C_{B\bar B} \,(\text{TeV}^2)$ & $C_7+0.24 C_8$ & $\mu_{\gamma\gamma}$\\ 
\hline
BP1 & 750.8 & 280.0 & 244.9 & 33.61 & $-1.03$ & 0.07$\times10^{-5}$ & $-0.006$ & 0.98\\
\hline
BP2 & 826.8 & 290.0 & 260.0 & 17.44 & $-0.90$ & 0.05$\times10^{-5}$ & $-0.005$ & 0.98\\
\hline
BP3 & 902.02 & 300.0 & 270.0 & 9.8  & $-0.79$ & 0.04$\times10^{-5}$ & $-0.004$ & 0.98\\
\hline
BP4 & 1001.8 & 320.0 & 282.8 & 5.0  & $-0.67$ & 0.03$\times10^{-5}$ & $-0.003$ & 0.98\\
\hline
\end{tabular}
}
\caption{The benchmark points and the production cross
sections at 14~TeV LHC are shown in columns (2--5). The
corresponding Wilson coefficients (Eqs.~(\ref{eqn:CBB})) are
given in columns (6--8). The Higgs boson signal strength is also provided in the last column.}
\label{bp}
\end{table}

\begin{table}[!h]
\centering
\begin{tabular}{|c|c|c|c|c|c|}
\hline
\qquad$h_1$\qquad\qquad & \qquad$h_2$\qquad\qquad & \qquad$h_3$\qquad\qquad &  \qquad$h'_1$\qquad\qquad & \qquad$h'_2$\qquad\qquad & \qquad$h'_3$\qquad\qquad  \\ 
\hline
0.01 & $-0.01$ & 0.52 & 5.0$\times10^{-5}$ & 3.0 & 0.01\\
\hline
\end{tabular}
\caption{The values of the Yukawa-type couplings kept fixed
for all the benchmark points. }
\label{tab:yuk}
\end{table}

The standard search for scalar leptoquarks at the collider experiments are
carried out by looking at the $2\ell+2j$ channels\,\cite{Okumura:2021yjv,ATLAS:2020dsk,D0:2006wsn,ATLAS:2016wab}.
These searches generally assume that the leptoquarks couple to a SM lepton and a SM quark. For such an assumption 
the leptoquark can be produced singly and in a pair and the decay of
the leptoquark will give no missing transverse momenta when the decay contains a charged SM lepton. As expected, 
selection cuts on $\mET$ are usually kept low for such scenarios. On the other hand, in our model, the leptoquark couples 
to a SM quark and a BSM VLL, owing to the discrete $\mathbb{Z}_2$ parity. This BSM lepton then decays to a 
scalar DM and a SM lepton. As a result, in our model, one expects
large $\mET$ from the decay of leptoquark. Therefore, the current searches for leptoquark at the LHC do not put
very strong constraints on our model parameters. In some cases, the leptoquark has been studied in the $2\ell+2t$
channel\,\cite{Okumura:2021yjv,ATLAS:2021oiz,CMS:2018qqq},
where one expects a little more $\mET$, if the top quark decays
leptonically. However, these searches also do not constrain the model parameters of our model significantly.

Interestingly, the search for the leptoquark in this particular channel ($2\mu+2b+\mET$) has been carried out by the ATLAS and 
CMS collaborations at the 13~TeV LHC run\,\cite{ATLAS:2016xcm,ATLAS:2016lsr,ATLAS:2017drc,CMS:2017fij}.
The strongest bound comes from the search for top squark pair production in the same channel\,\cite{ATLAS:2016xcm}
and hence this analysis when recasted for our model, could constrain our model parameters as well.
We have scanned our model parameters using the recasting tool, i.e., the {\tt CheckMATE}\,\cite{Drees:2013wra} package 
and have realized that leptoquarks, heavier than 750~GeV are not excluded by the above search at 95\% C.L. We have 
also checked that all the benchmark points are allowed by the above search.

Since we demand that the BSM particles are odd under $\mathbb{Z}_2$ symmetry, the lightest particle may become the
DM candidate. In the above four benchmark points, the real scalar $S$ is the lightest $\mathbb{Z}_2$ odd particle and
hence it plays the role of DM candidate. To study the DM phenomenology, the {\tt CALCHEP}\,\cite{Belyaev:2012qa} compatible 
model files are obtained from {\tt SARAH}\,\cite{Staub:2008uz} and then included in {\tt MicrOMEGAs}\,\cite{Belanger:2014vza}, which calculates the DM observables like relic density $\Omega_{\rm DM} h^2$, spin-dependent ($\sigma_{\rm SD}$) and 
spin-independent ($\sigma_{\rm SI}$) cross sections, and the thermally averaged annihilation 
cross sections ($\langle \sigma v \rangle)$. As the scalar DM does not couple with the nucleons, the direct detection 
constraints are easily satisfied in our model. We also find that our benchmark points are compatible with 
relic abundance obtained from the {\tt PLANCK} experiment\,\cite{Planck:2018vyg}. 
For our four benchmark points, BP1, BP2, BP3 and BP4, the relic density is $7.79 \times10^{-3}, 4.16 \times10^{-3},
3.98 \times10^{-3}$ and $7.28 \times10^{-3}$ respectively. This suggests that the DM is under-abundant for these benchmark 
points. However, tuning the Yukawa coupling ($h'$) parameters could give us the correct relic abundance. Since we 
are interested in the leptoquark signal at LHC, we postpone the discussion about dark matter for separate work.

For the LHC analysis, we implement the model file in
{\tt SARAH}\,\cite{Staub:2008uz} and generate the required UFO
which is used to generate events in
{\tt MADGRAPH}\,\cite{Alwall:2014hca}. The spectrum files for the benchmark points are generated using
{\tt SPheno}\,\cite{Porod:2011nf}. The events generated in {\tt MADGRAPH}
are then passed on to {\tt PYTHIA8}\,\cite{Sjostrand:2014zea} for
showering and hadronization. The detector simulation is done
in {\tt DELPHES}\,\cite{deFavereau:2013fsa} using the default CMS card, where the jets are constructed using 
the anti-$K_T$ algorithm with a jet formation radius of $R = 0.4$. For the SM background processes with hard jets, 
proper MLM matching scheme\,\cite{Hoeche:2005vzu} has been applied. The charged leptons in the final state are isolated 
by choosing  $\Delta R_{\ell i} > 0.4$, where $i$ represents either a jet or a lepton. For generating signal and background 
events at the parton level, we use the following kinematic acceptance cuts on the partons and leptons (electron and muon):
\begin{eqnarray}
 p_T(j,b) &>& 20 ~{\rm GeV}\,; \quad |\eta(j)| < 4.7 \,; \quad |\eta(b)| < 2.5 \,, \nonumber \\ 
 p_T(\ell) &>& 10 ~{\rm GeV}\,, \quad |\eta(\ell)| <2.5 \,.
 \label{basic_cut}
\end{eqnarray}

\begin{figure}[!h]{\centering
		\subfigure[]{\includegraphics[height = 5.5 cm, width = 8 cm]{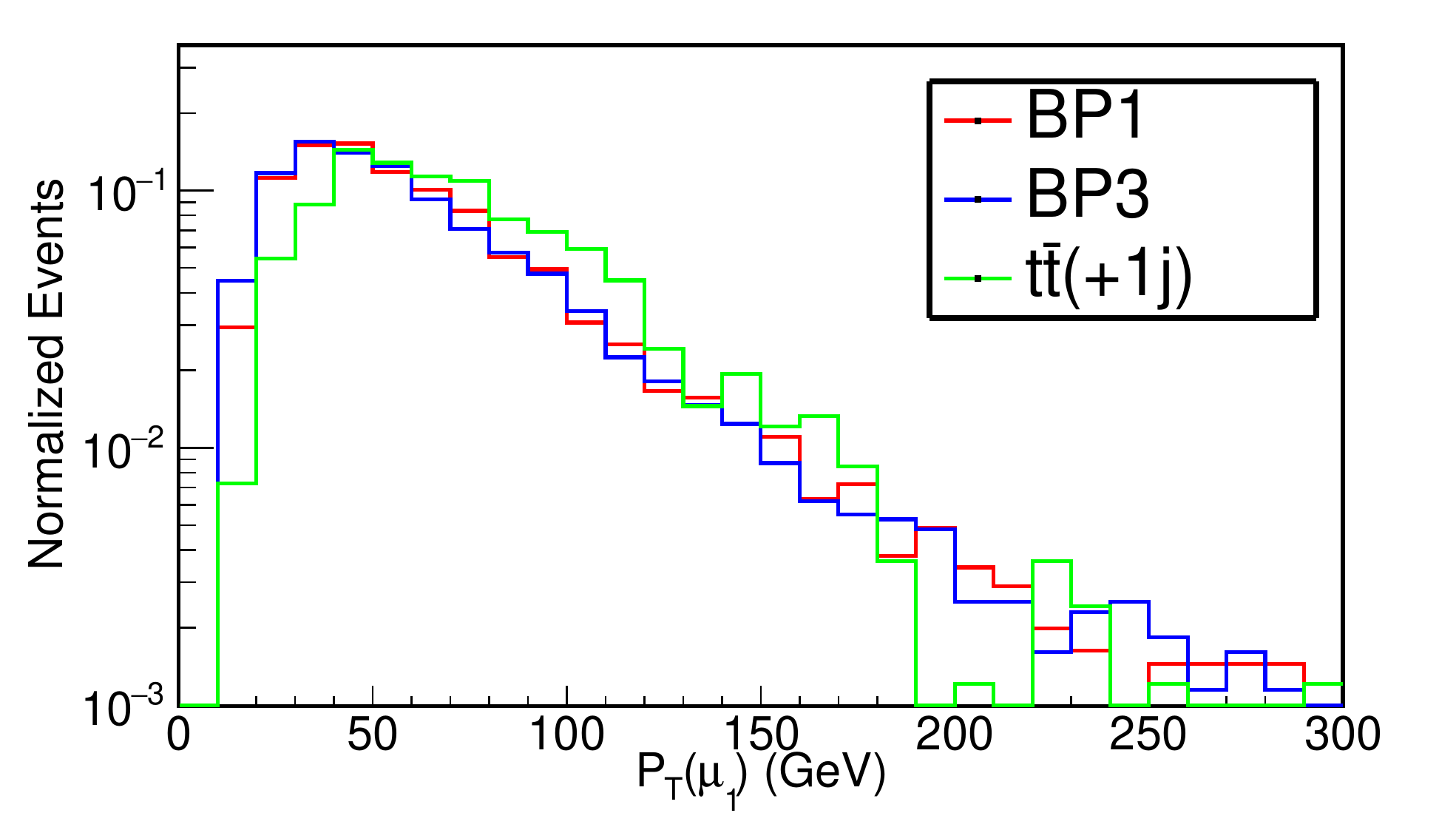}} 
		\subfigure[]{\includegraphics[height = 5.7 cm, width = 8 cm]{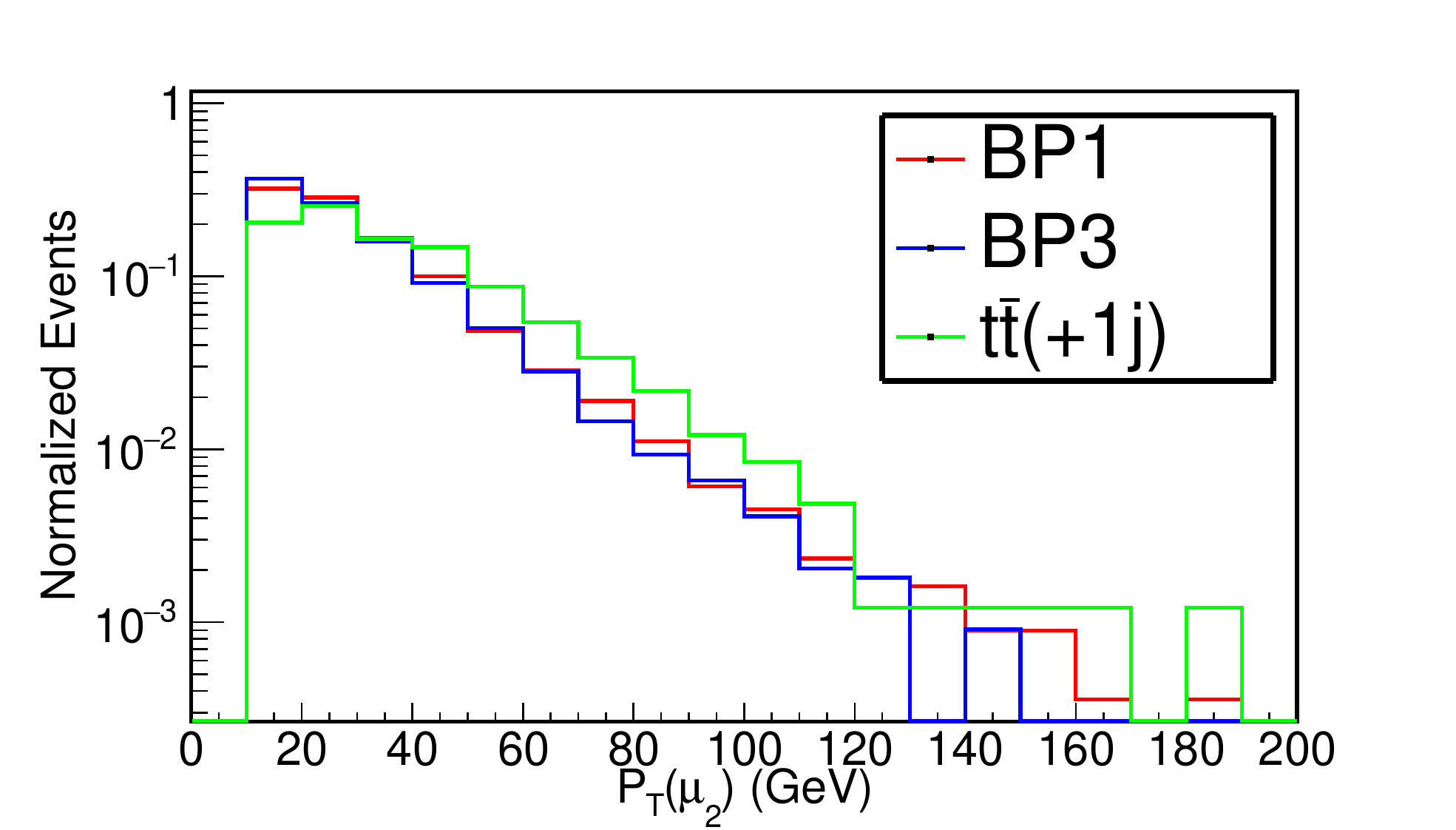}}
		\subfigure[]{
			\includegraphics[height = 5.5 cm, width = 8 cm]{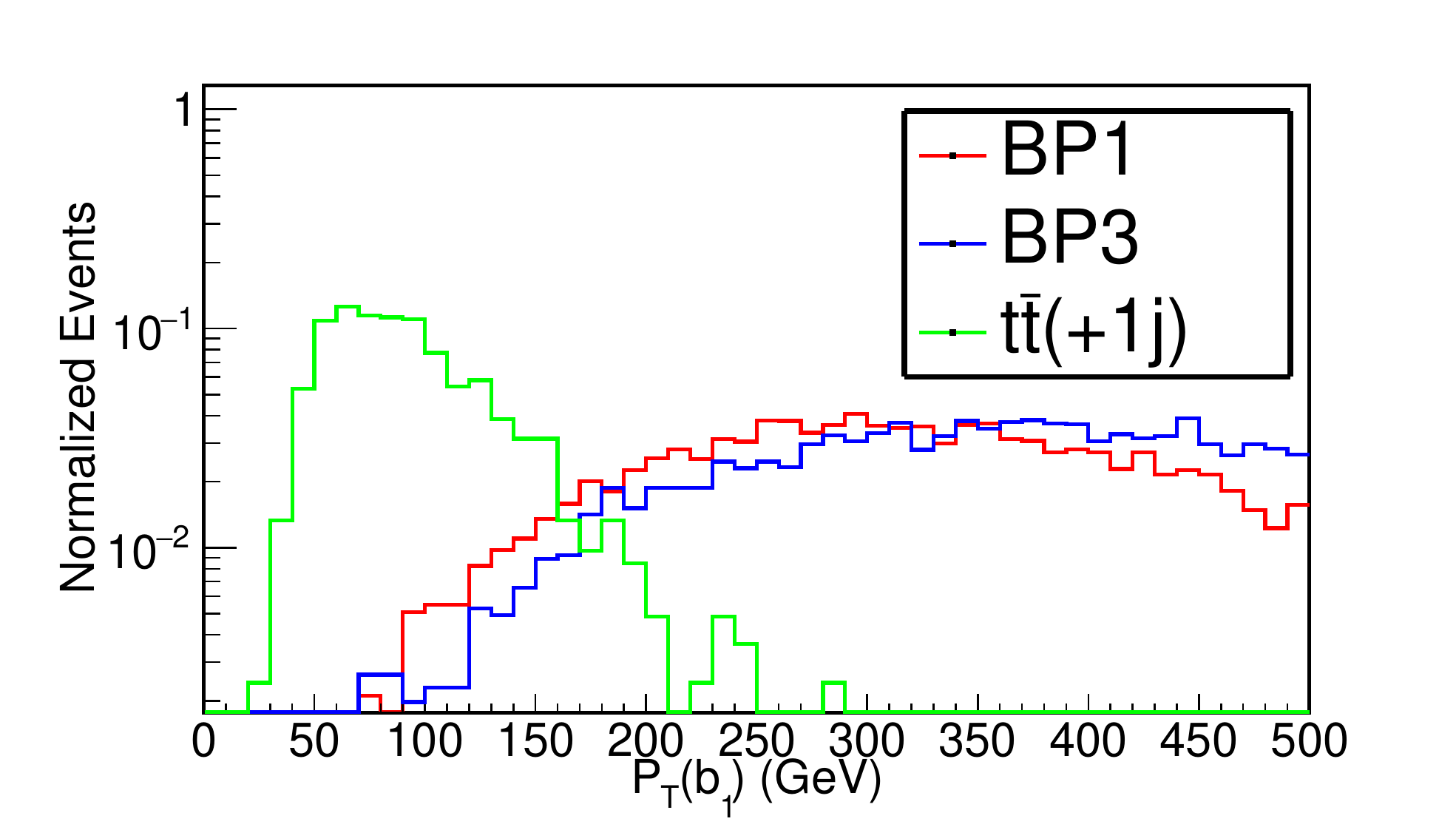}} 
		\subfigure[]{
			\includegraphics[height = 5.5 cm, width = 8 cm]{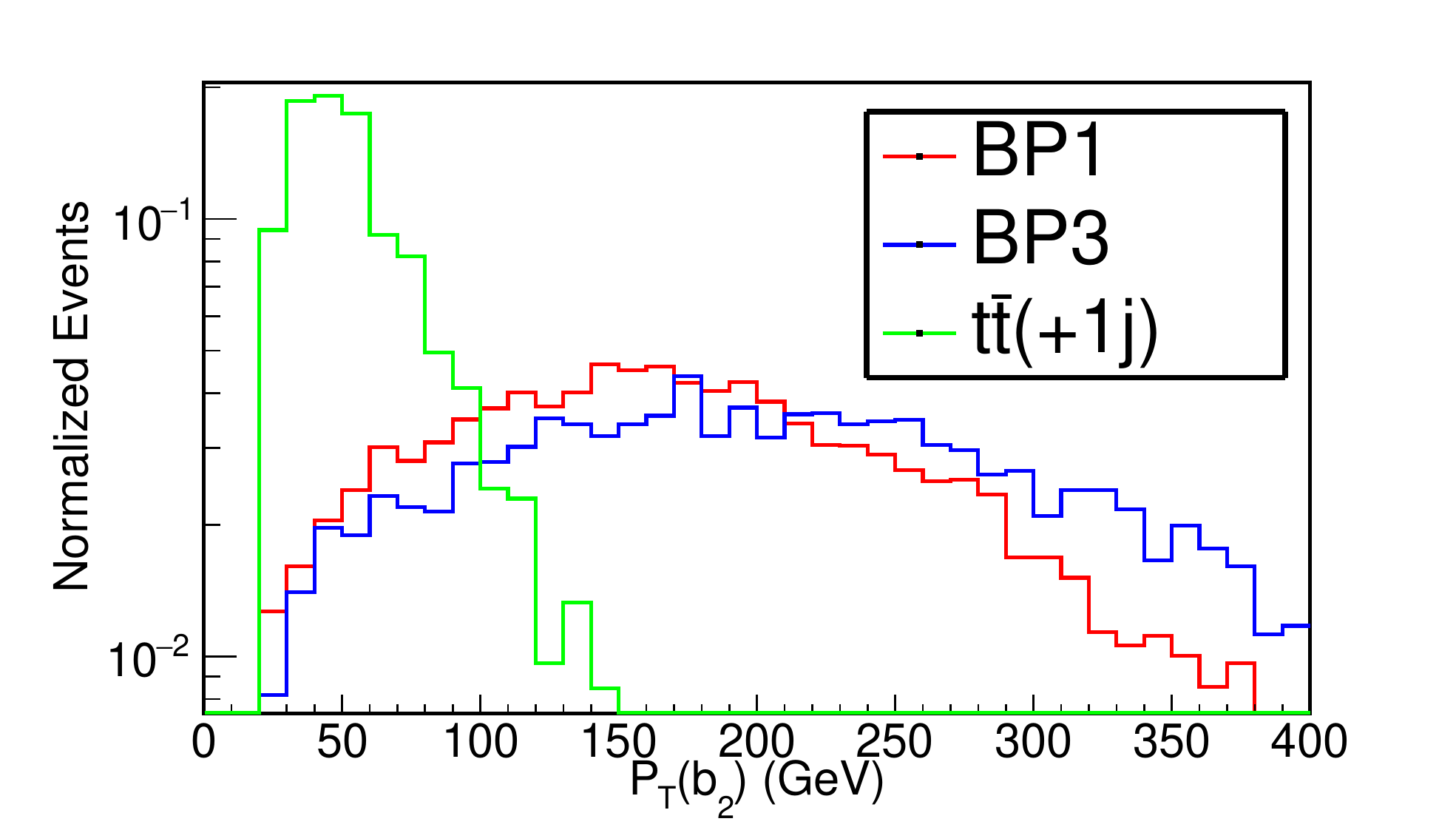}} 
		\subfigure[]{
			\includegraphics[height = 5.5 cm, width = 8 cm]{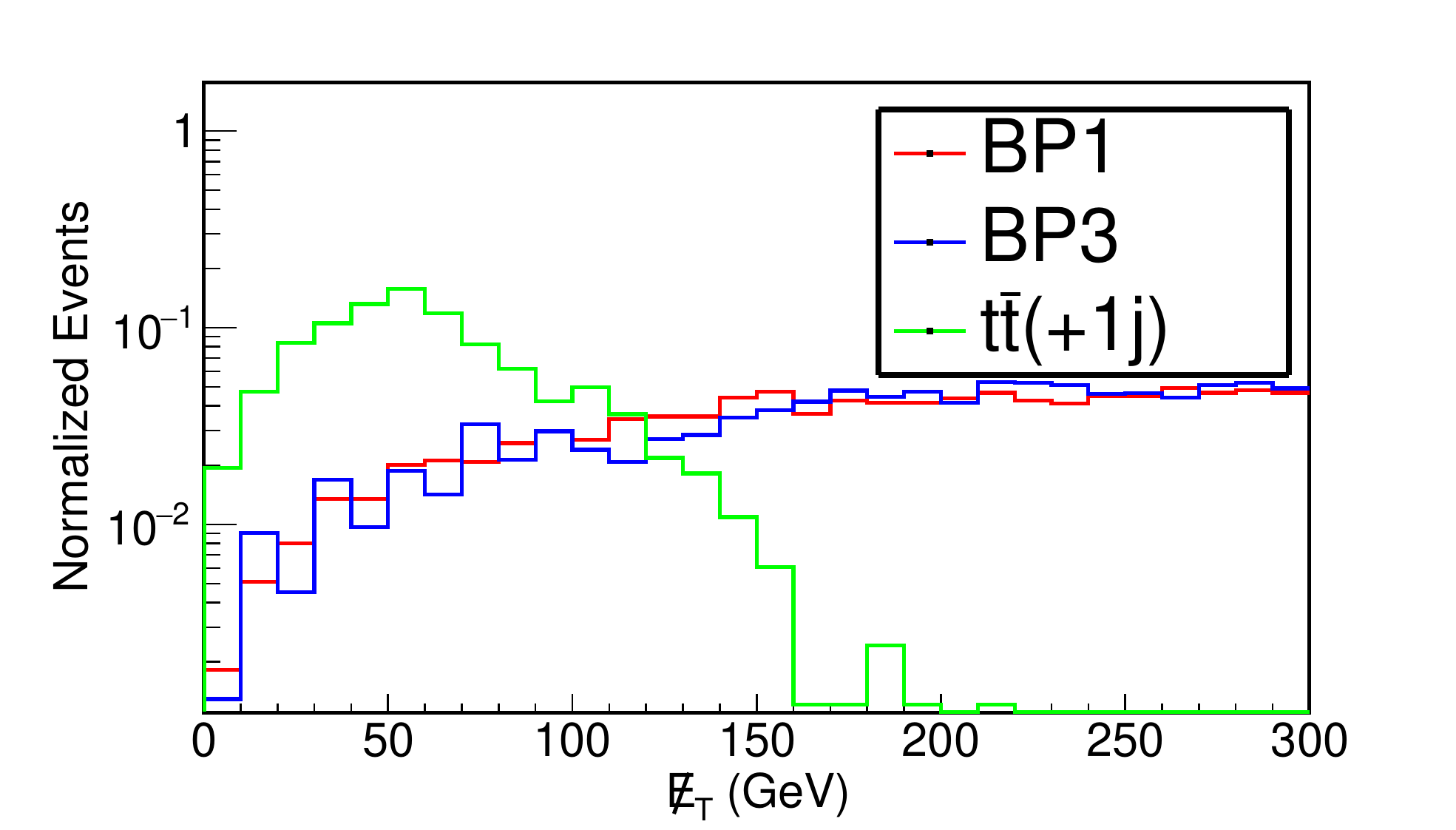}} 
	}
	\caption{Normalized distribution of the kinematic variables
		for the signal and dominant SM background. }
	\label{kinematics}
\end{figure}
The $b$-jets are tagged using a $p_T$ dependent efficiency given as 
\begin{equation}
\epsilon_b = \tt 0.85\,\tanh(0.0025\,p_T)\,\dfrac{25.0}{(1+0.063\,p_T)}. \nonumber
\end{equation}
Similarly a mistagging efficiency for $c$-jets being wrongly identified as $b$ jets, given by
\begin{equation}
\epsilon_{c \to b} = \tt 0.25\,\tanh(0.018\,p_T)\,\dfrac{1}{(1+ 0.0013\,p_T)}, \nonumber
\end{equation} 
is included and a mistag efficiency of $\epsilon_{l \to b} = \tt (0.01+0.000038\,p_T)$ for 
the light jets are also included.

We now provide the details of our cut-based analysis of the signal and SM background events, which
maximize the signal to background ratio. Along with the basic cuts mentioned in Eq.~\ref{basic_cut}, we put 
additional selection cuts on the following kinematic variables as described below:
\begin{itemize}
    \item $\bf{p_T(\mu)}$: We depict the $p_T$ distribution
    of the leading and sub-leading muon in
    Figs.~\ref{kinematics}(a) and \ref{kinematics}(b),
    respectively. As our signal contains two muons, we
    select one leading and one sub-leading muon with
    transverse momentum $p_T > 10$ GeV and reject events
    containing more than two muons. We can see that the
    signal and background events peak in the same $p_T$
    range. We find that the muon in the signal events have a $p_T$ cut-off 
    dependent on the mass of the VLL. Therefore a lower cut on the $p_T$ does not help 
    to reduce the background. Instead, an upper cut on the
    leading muon $p_T(\mu_1) < 180$~GeV is helpful in removing the SM tail 
    and improving the signal sensitivity.
    
    \item $\bf{p_T(b)}$: We show the normalized $p_T$
    distribution for the leading and sub-leading $b$-jet in
    Figs.~\ref{kinematics}(c) and \ref{kinematics}(d),
    respectively. To ensure that our signal contains only
    two $b$-jets, we reject events with a third $b$-jet with
    $p_T > 20$~GeV. For the background, the $b$-jets come
    from the decay of top quark, whereas for the signal the
    $b$-jets come directly from much heavier leptoquarks. So the signal 
    distribution peaks at a higher value compared to the SM background. We put a 
    lower cut on the transverse momentum $p_T(b_1) > 200$ GeV and
    $p_T(b_1) > 100$~GeV to enhance the signal over
    background. This cut also helps in drastically suppressing the  
    $t\bar{t}V$ and di-boson+dijet backgrounds, finally leaving us with background events 
    coming dominantly from the inclusive $t\bar{t}(+1 jet)$ production in SM.
    
    \item $\bf{\mET}$: The $\mET$ distribution is plotted in
    Fig.~\ref{kinematics}(e). As our signal contains a heavy
    dark matter scalar candidate $S$, the $\mET$ peaks at a
    higher value in contrast to the background where the
    $\mET$ comes only from the neutrinos arising from the $W^{\pm}$ decay. We 
    put a lower cut on missing transverse energy as $\mET > 200$ GeV
    to suppress the SM background further.
\end{itemize}
The detailed outcome of our cut-flow choice for different benchmark points is shown in Table~\ref{cutflow}.

\begin{table}[htbp!]
	\centering
	\begin{tabular}{|p{3.0cm}|c|c|c|p{3.0cm}|}
		\cline{2-4}
		\multicolumn{1}{c|}{}& \multicolumn{3}{|c|}{Number of Events after cuts ($\mathcal{L}=3000$ fb$^{-1}$)} & \multicolumn{1}{c}{} \\ \cline{1-4}
		SM-background  
		 & $p_T(\mu_1)$ cut  &  $p_T(b)$ cut    &  $\mET$ cut   & \multicolumn{1}{c}{}
		\\ \cline{1-4} 
                             $t\bar{t}(+1jet)$  & 974682 & 8361  &  717 \\ \cline{1-4} 
                             $t\bar{t}V$ & 585 & 32 & 5 \\ \cline{1-4} 
                            $VVjj+VVb\bar b$ & 9223 &  41 &  15 \\ \cline{1-4} 
                             Total background & 984490 & 8434 & 737 \\ \cline{1-4} \hline
		
			\multicolumn{1}{|c|}{Signal }  &\multicolumn{3}{|c|}{} &\multicolumn{1}{c|}{Significance reach at 3000 fb$^{-1}$}  \\ \cline{1-5}
		\multicolumn{1}{|c|}{BP1} &  546  &  318 & 318  &\multicolumn{1}{|c|}{11.0$\sigma$}  \\ \hline
		\multicolumn{1}{|c|}{BP2} & 244   & 199  & 160  &\multicolumn{1}{|c|}{5.7$\sigma$}  \\ \hline
		\multicolumn{1}{|c|}{BP3} &   125 &  105 & 88  &\multicolumn{1}{|c|}{3.2$\sigma$}  \\ \hline
		\multicolumn{1}{|c|}{BP4} & 57   & 51 &  45 &
		 \multicolumn{1}{|c|}{1.6$\sigma$}  \\ \hline 
	\end{tabular}
\caption{The cut-flow for signal and backgrounds for $2\mu
+ 2b + \mET$ channel along with the significance for
benchmarks BP1, BP2, BP3 and BP4 to be probed at 14~TeV
LHC with 3000 fb$^{-1}$ luminosity.}
\label{cutflow}
\end{table}

The signal significance is then calculated by using the
formula\,\cite{Cowan:2010js}
\begin{equation}
\mathcal{S} = \sqrt{2\Big[(S + B) \log\Big(\frac{S + B}{B}\Big)- S\Big]},
\end{equation}
where $S (B)$ represents the number of signal (background)
events surviving after all the cuts are applied. It can be seen from Table~\ref{cutflow} that, as 
the leptoquark mass increases, the signal significance decreases Although the cut efficiencies are more useful 
for the signal coming from the heavier states, the production cross section falls as shown in Table~\ref{bp}. 
We find that it will be possible to probe the benchmark points BP1, BP2, and, BP3 
with statistical significance of $11.0\sigma, 5.7\sigma, 3.2\sigma$ respectively, while BP4 with the 
leptoquark mass of 1 TeV gives a $1.6\sigma$ statistical significance with 3000 fb$^{-1}$ 
integrated luminosity.

\section{Conclusion}~\label{conc}
The observation of muon and flavor anomalies has generated a lot of renewed interest in new ideas 
of BSM physics in the particle physics community. The extension of SM by new fields usually provides 
contributions to these anomalies via higher-order loops. 

In this work, we have studied a model which extends the SM by a scalar leptoquark, one generation of VLL and one SM
singlet scalar. The new fields are odd under a $\mathbb{Z}_2$ symmetry. By virtue of this $\mathbb{Z}_2$
symmetry the scalar or the lightest neutral VLL acts as a DM candidate. Any mixing with the SM fields is also avoided
due to this discrete parity being unbroken in the model. The new fields couple to the SM particles via Yukawa-type
interactions. The VLL and the scalar contribute to the anomalous magnetic moment of muon at the one-loop. We see that
for a significant range of the parameter space, this anomaly can be satisfied within the $3\sigma$ error bar of the current
data. The Yukawa interaction also provides an extra contribution to the LFV. We check for parameter space that
is allowed by all such LFV constraints and QFV constraints, namely $b\to s \mu\mu$, $b\to s\gamma$ and 
$B_s-\bar{B_s}$ mixing. 

We then chose representative benchmark points for leptoquark mass and look for a distinct signal in
$2\mu+2b+\mET$ final states which can be probed at the 14~TeV LHC run. We also note that the chosen benchmark points
satisfy DM constraints as well as account for the measured muon anomaly. Based on a simple cut-based analysis we 
find that most of the benchmark points lead to signals with more than $3\sigma$ significance at 3000 fb$^{-1}$
integrated luminosity. We also conclude that a scalar leptoquark with masses above $\gtrsim$ 1 TeV would be difficult 
to observe in the simple-minded cut-based analysis and may require more sophisticated machine-learning methods to 
have any chance of observation. Our analysis also shows that current limits on the scalar leptoquarks which decay
directly to SM particles would become much weaker in the presence of non-standard decays of the leptoquark. This can
be a highly probable scenario if one considers explaining the flavor and muon anomalies in a common setup and looks at
the collider signatures of such a model. 

\section{Acknowledgement}
NG and SKR would like to thank the Regional Centre for
Accelerator-based Particle Physics (RECAPP), Harish-Chandra
Research Institute, Department of Atomic Energy, Government
of India for financial support. TS would like to acknowledge
RECAPP for providing hospitality when this work was ongoing. 


\providecommand{\href}[2]{#2}\begingroup\raggedright\endgroup


\end{document}